\documentclass[groupedaddress,english, superscriptaddress]{revtex4-1}

\usepackage{soul}
\usepackage{color}
\usepackage{censor}
\usepackage{subfiles}
\usepackage{titlesec}
\usepackage[normalem]{ulem}
\usepackage{url}
\usepackage{mathtools}
\usepackage{graphicx}

\usepackage{amsmath,amsthm, amssymb}
\usepackage{appendix}

\definecolor{blue}{rgb}{0.75,0.75,0}
\newcommand{\mehrnaz}[1]{\textcolor{blue}{#1}}
\definecolor{green}{rgb}{0.,0.5,0.5}

\definecolor{dred}{rgb}{0.5,0,0}

\begin{document}

\title{Characterising human energy consumption}
\title{How far is the standard load profile from the reality?}
\title{Characterising human energy demand}
\title{Characterising fluctuations in human electric energy demand}
\title{Characterising the human  \mehrnaz{Holger's suggestion: residential instead of human? I agree with that and you?} electricity consumption}
\title{Data-Driven Load Profiles and the Dynamics of Residential Electric Power Consumption}

\author{Mehrnaz Anvari}
\thanks{contributed equally}
\affiliation{Potsdam Institute for Climate Impact Research (PIK), Member of the Leibniz Association, P.O. Box 60 12 03, D-14412 Potsdam, Germany}

\author{Elisavet Proedrou}
\thanks{contributed equally}
\affiliation{DLR Institute for Networked Energy Systems}

\author{Benjamin Sch\"afer}
\thanks{contributed equally}
\affiliation{School of Mathematical Sciences, Queen Mary University of London, United Kingdom}

\author{Christian Beck}
\affiliation{School of Mathematical Sciences, Queen Mary University of London, United Kingdom}

\author{Holger Kantz}
\affiliation{Max Planck Institute for the Physics of Complex Systems, D-01187 Dresden, Germany}

\author{Marc Timme}
\affiliation{Chair for Network Dynamics, Center for Advancing Electronics Dresden (cfaed) and Institute for Theoretical Physics, Technical University
of Dresden, 01062 Dresden, Germany}




\begin{abstract}

The dynamics of power consumption constitutes an essential building block for planning and operating energy systems based on renewable energy supply. Whereas variations in the dynamics of renewable energy generation are reasonably well studied, a deeper understanding of short and long term variations in consumption dynamics is still missing. Here, we analyse highly resolved residential 
electricity consumption data of Austrian and German households and propose a generally applicable methodology for extracting both the average demand profiles and the demand
fluctuations purely from time series data. The analysis reveals that demand fluctuations of individual households are skewed and consistently highly intermittent. We introduce a stochastic model to quantitatively capture such real-world fluctuations. The analysis indicates in how far the broadly used standard load profile (SLP) may be is insufficient to describe the key characteristics observed. These results offer a better understanding of demand dynamics, in particular its fluctuations, and provide general tools for disentangling mean demand and fluctuations for any given system. The insights on the demand dynamics may support planning and operating future-compliant (micro) grids in maintaining supply-demand balance.
\end{abstract}

\maketitle

\clearpage


\clearpage

\section{Introduction}
\label{sec:Introduction}
Electrical energy is an essential part of the daily life that should be generated, transmitted, stored and, finally, consumed. Generation and storage of electricity shall match the dynamic consumption of residential, industrial and other sectors at all times. To maintain the balance between the electricity generated by energy providers, and the electricity consumed by consumers, energy suppliers need to know the electricity required by all consumer sectors on a broad range of time scales, i.e. seconds to days. Estimating the typical variations in the electricity demand over the course of a day yields a load profile, which can be attained either through the definition of a methodology to extract a load profile from empirical data, the creation of a model or a combination of both.\\

Research aiming at developing load profiles goes back to at least the 1940s \cite{GrandJeanetal2012}, however the issue of finding a precise, high resolution load profile is becoming more and more urgent due to increasing population, electrical heating systems, electrical vehicles for transportation, solar home systems as well as the increasing share of fluctuating renewable energy (RE) feed-in and the construction of distributed power grids, especially smart grids. In 1999, the first methodologically systematic German household load profile, known as the H0 Standard Load Profile (H0 SLP), was developed \cite{SLP} and has since been in use without alterations in at least Germany and Austria \cite{AustriaSLP}. \\

We focus here on the residential sector consuming around 29\% of all electricity in the European Union \cite{EuropeanEnviromentEnergy2018}. Although initially household load profiles used a temporal resolution of around one hour, newer models have a temporal resolutions up to and including one second, due to the recent availability of highly resolved data sets of electricity consumption and the usage of smart meters in selected houses (see Supplementary Note 1). These new data sets allow the grid operators to record the electricity consumption of individual houses at high temporal resolutions. Several previous studies \cite{GREEND,Wright2007,MarszalPomianowskaetal2016} analysed high temporal resolution datasets and reported the presence of extreme and significant peaks in the datasets which have not been reported for data sets with temporal resolutions of 15 minutes to 1 hour. 
However, a structured approach to translate these high-resolution data sets into usable load profiles is not readily available to date. Therefore, in this work we first analyse highly resolved electricity consumption data for groups of houses in Austria and Germany. Our data-driven analysis indicates the potential for the presence of strong fluctuations and high levels of unpredictability in the distribution grids, see Section~\ref{sec:Necessity_of_new_demand_models}. Then, we use the same analysis to introduce the new load profile which is consistent with high-resolution electricity consumption data.\\


As mentioned above, one important reason of having a high resolution load profile is the increasing share of RE feed-in. In contrast to electric energy generated from burning fossil fuels, RE feed-in is weather-dependent, intermittent, and highly variable \cite{liu2011passivity,Milan2013,Anvari2016}. It thus becomes harder to balance supply and demand. As the share of renewable feed-in is increasing, recent and current research focuses on gaining a deeper understanding of the electricity generated by RE as well as advanced approaches of balancing demand and supply, e.g. by load-shifting \cite{lopez2015demand,logenthiran2012demand}. In contrast, the dynamics of electricity consumption is far from understood, in particular on the residential level, partially because electricity consumption data are difficult to obtain.\\

Here, we analyse the dynamic characteristics of the residential power consumption at a high temporal resolution and develop a generally applicable methodology to extract both the consumption trend and the consumption fluctuations. After reviewing the need for a new demand model in more technical detail (Section \ref{sec:Necessity_of_new_demand_models}), we analyse the electricity consumption data of Germany and Austrian houses measured for several weeks. We disentangle the variations in the consumption dynamics into two main factors, the average load profile (ALP) and the statistics of short-time fluctuations of the ALP. First, through the application of the empirical mode decomposition (EMD) \cite{EEMD}, we extract the average load profile from the time series data. This extracted trend captures the demand much more accurately than the often-used H0 SLP (Section \ref{sec:Demand_trend}). Next, we use superstatistics to model the fluctuations around this trend into a stochastic fluctuation profile (SFP). Combining the trend and fluctuation analysis, we successfully reproduce a synthetic high resolution load profile, yielding a full data-driven load profile (DLP). Our modelling approach can be readily generalised for use with other data sets and in order to facilitate this we are providing executable code as (Section \ref{sec:Demand_fluctuations}). We thereby develop a load profile methodology that is applicable to existing power grids and data sets, and also provides the tools for extracting load profiles in different regions and under different boundary conditions, for instance for microgrids with high shares of on-site generation or electrical cars.\\

\section{Complex demand dynamics -- the necessity of new load profiles}
\label{sec:Necessity_of_new_demand_models}
\begin{figure}[ht]
\centering
\includegraphics[width=0.9\columnwidth]{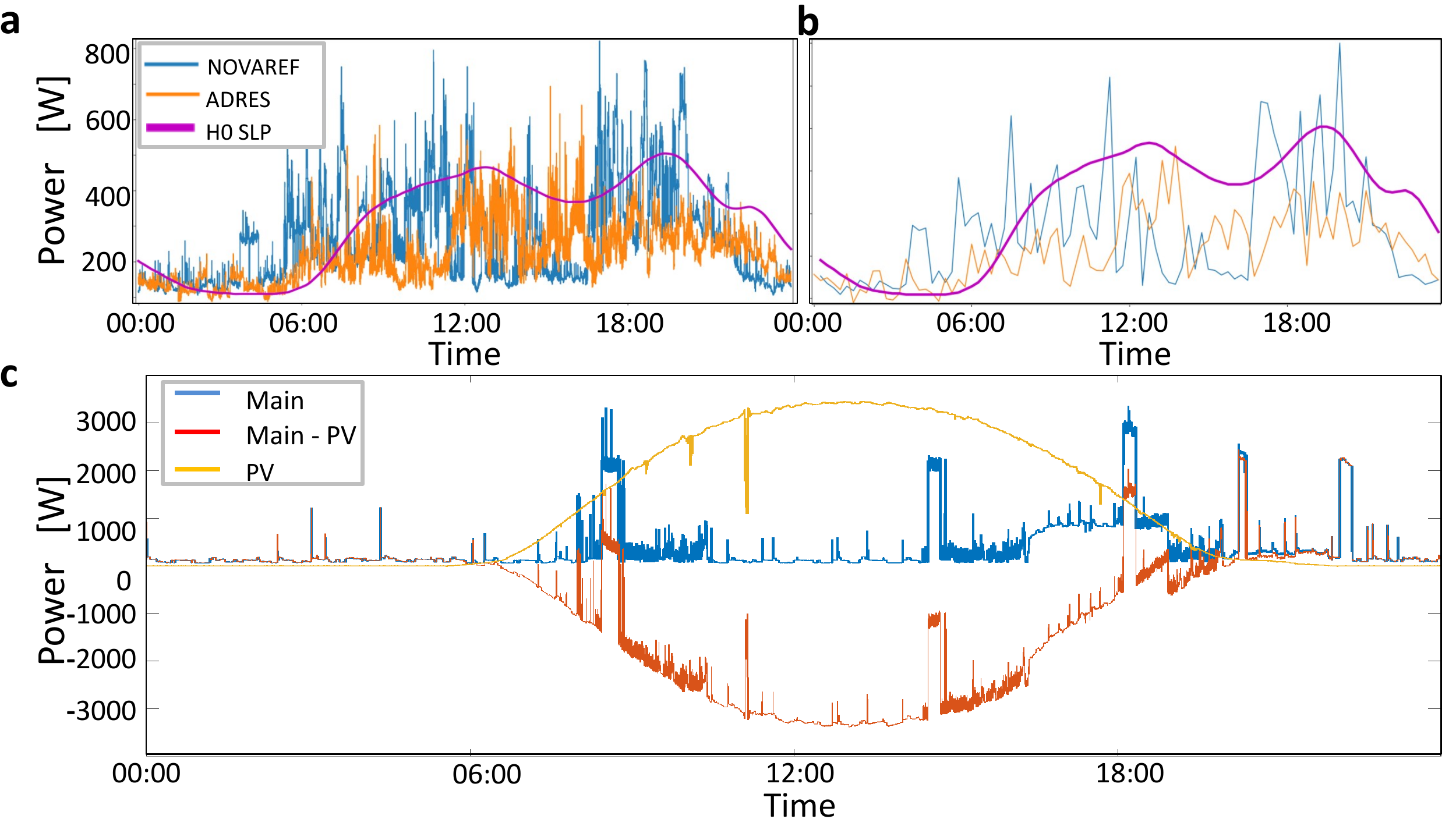}
\caption{
{\bf Systematic, asymmetric and intermittent deviations of empirical power consumption dynamics and German standard load profile (H0 SLP)} (a) We compare the H0 SLP with the averaged real consumption data of a single day in winter for the NOVAREF (12 houses) and  ADRES (30 houses) data sets. The H0 SLP not only fails to capture the correct daily trend, but we also observe large fluctuations of the consumption at short time scales. (b) Shows the same data sets at 15 minutes resolution to emulate the temporal resolution that was used to generate the H0 SLP. Here the failure of the H0 SLP to capture the correct daily trend is even more pronounced. (c) Shows the recorded household power consumption with and without photovoltaics as well as the electricity generated by the PV module of a single household. Note the smaller spikes visible in panels (a) and (b) because the data is averaged over 12 and 30 houses, respectively, for NOVAREF and ADRES, while in panel (c) only a single household is shown.}
\label{fig:SLP_vs_RealData}
\end{figure}

\begin{figure}[ht]
\centering
\includegraphics[width=0.9\columnwidth]{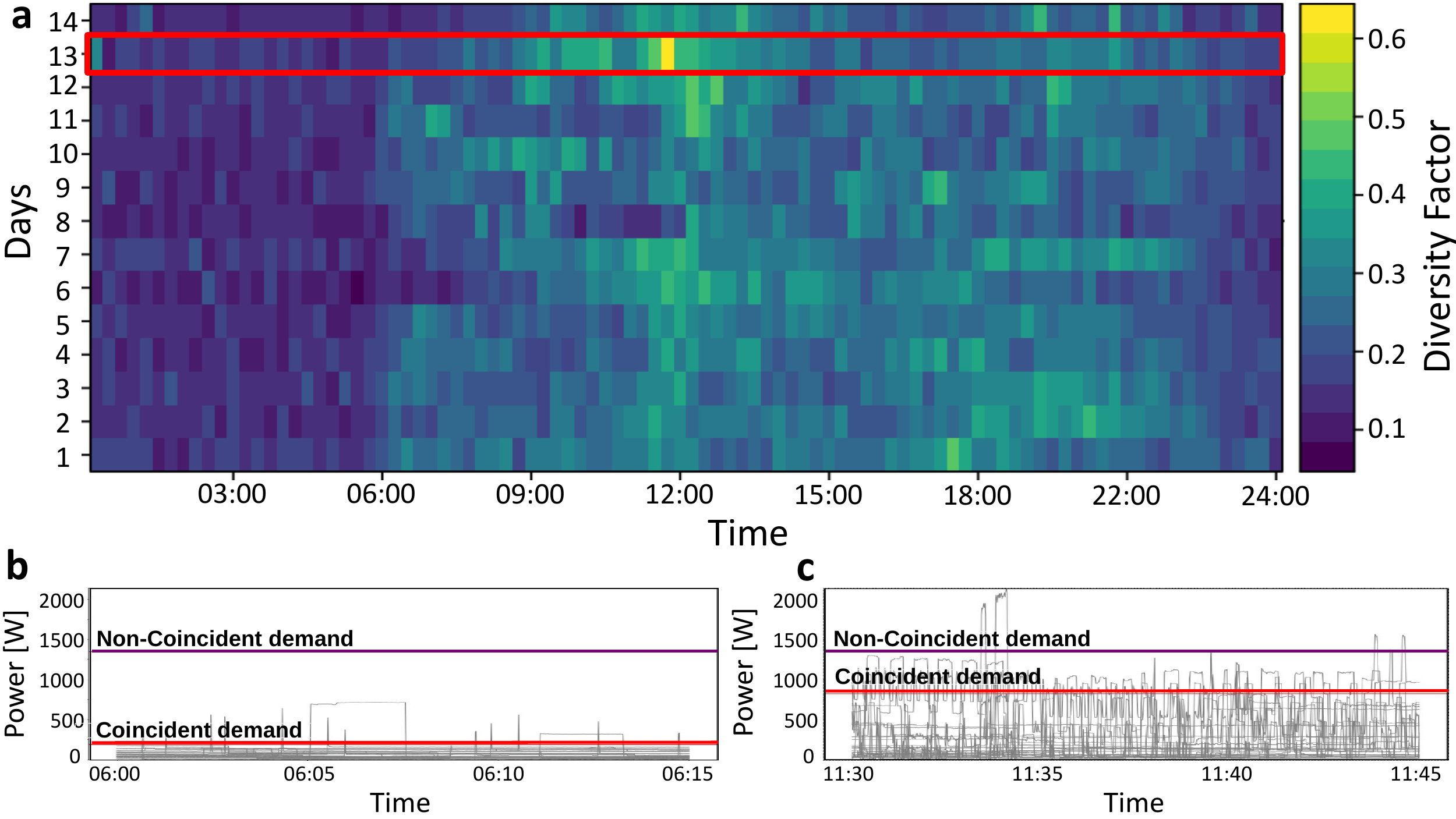}
\caption{{\bf Indicating the interaction between households energy consumption by considering the diversity factor} (a) shows the diversity factor or the ratio between coincident demand, $P_{cd}$, and non-coincident demand, $P_{ncd}$, of 30 households belonging to the ADRES data set every 15 minutes and for 14 days. As it is clear, there are some time intervals during a day whose diversity factor reaches 0.6. (b) and (c) respectively show the energy consumption trajectory of all 30 households from 6:00 to 6:15 and 11:30 to 11:45. It is clear from the figures that the energy consumption of all houses is low around the 6:00 and then is gradually increasing together around 11:00 and, consequently the difference between $P_{cd}$ and $P_{ncd}$ is decreasing proving the obvious interaction between houses consuming the electricity.}
\label{fig:divercity_factor}
\end{figure}

Notwithstanding recent advances, energy suppliers still mostly use the older load profiles, such as the H0 standard load profile, which only has a 15 minute temporal resolution. In Germany, the standard load profile (H0 SLP), was introduced in 1999.
Ninety percent of the residential load data used in its creation were measured in the 1970s or earlier with an hourly temporal resolution. Only 10\% of the measured load data had a temporal resolution of 15 minutes. Here, we review some of the recent advances towards a modern load profile and touch on the persistent need for a generally applicable consumption framework which we provide in the next sections.\\
\newline
Three well-known model classes exist to describe a household load profile. {\it Top-down} or conditional demand analysis models are downward models that use the total electricity consumption estimates of multiple households as well as macro-variables to predict the household energy consumption and generate household load profiles \cite{Parti1980,Aigner1984}. {\it Bottom-up models} use micro-variables as input, such as the number of active occupants, the appliances' energy demand and usage time etc. They also often use Markov chains to generate household load profiles \cite{Capassoetal1994,PaateroLund2006}. Finally, {\it hybrid} models employ a combination of the techniques used in top-down and bottom-up models to build up a Statistical Adjusted Engineering (SAE) model \cite{Train1985}. A detailed analysis of how these models have been applied was recently reviewed in \cite{Estheretal2016}. At present, many demand side management models exist that can generate daily residential electricity load profiles, see Supplementary Note 1, \cite{BethLoadReview} and \cite{GrandJeanetal2012} for details. Only a few of these models use a high temporal resolution of the order of seconds and those require a lot of micro-parameters, which still leaves us with the need for an accurate, high-resolution, easy-to-use load profile to be developed.\\
\newline
A focus on higher temporal resolution is necessary to fully understand modern consumption patterns and respond quickly, for instance the disturbances caused by input fluctuations or regulatory or trading anomalies \cite{schafer2018non}. Recent statistical research on power consumption (see for example \cite{Wright2007} and \cite{ MarszalPomianowskaetal2016}) demonstrates that substantial differences exist between the statistical features of the highly resolved power consumption and consumption on a 15 minute time scale. Analysing power consumption on the short time scale of seconds to one minute reveals extreme consumption spikes, which are completely ignored in the 15 minute load profile \cite{Wright2007}. In Fig.~\ref{fig:SLP_vs_RealData}(a) and (b) a comparison of the H0 SLP with the load profiles of two residential data sets of high temporal resolution, measured in Germany and Austria is shown. The Austrian data set was recorded in 2009-2010 during the ADRES project \cite{ADRESReport}. The German data set was recorded between 2013 and 2016 during the NOVAREF project \cite{NOVAREFReport}. In the Supplementary Note 2 we report further analysis on data sets related to 70 households, recorded in August and September 2019 in Germany during the ENERA project.\\
\newline
As can be seen in Fig.~\ref{fig:SLP_vs_RealData}(a) and (b), the averaged single day load profiles of both the ADRES and the NOVAREF data sets, strongly deviate from the H0 SLP. Specifically, the trend (mean) of the load profile of the measured data is not well described by the H0 SLP and the fluctuations on short times scales are significant (see Section \ref{sec:Demand_trend}). Furthermore, the analysis of the statistics of the highly resolved power consumption reveals the presence of non-Gaussian, intermittent and asymmetric fluctuations, which need to be taken into account when designing demand side management and control mechanisms (see Section \ref{sec:Demand_fluctuations} and Supplementary Note 4.)\\
\newline
Some of the observed stochastic properties could be explained by the increased usage of new power generation and consumption devices: Rooftop photovoltaic panels (PVs) in the first case and new electronic devices such as electrical heating systems, smartphones, tablets, robot vacuum cleaners, electric vehicles etc. in the second case \cite{Destatis2018}. The usage of such devices will influence and alter the shape of the current standard load profile. As shown in Fig.~\ref{fig:SLP_vs_RealData}(c), the consumption can even reach negative values if houses are directly connected to local RE resources \cite{Braunetal2009}. Local fluctuations in solar and wind generation may further lead to coincident load profile spikes, both positive and negative. Therefore, the precise prediction and management of the power consumption of households on a short time scale is essential to ensure the stability of distributed grids. In addition to installing more on-site generations in the residential sector, replacing fossil-fuelled cars with electric ones, which are charged using household electricity, results in changes in the shape of the load profile of the households (see Supplementary Note 2), especially when organising consumers in micro grids \cite{Muratori2018}.
Since we expect an increasing number of such electrical cars, as well as an increasing penetration of strongly fluctuating RE generating electricity, a new data driven load profile based on modern and highly resolved measurements is urgently needed.\\
\newline
As mentioned above, extreme consumption spikes are completely ignored in the 15 minute load profile as they are averaged out. However, these spikes are of particular importance to lower voltage distribution grids, where coincident consumption can dominate the consumption patterns locally or on a country scale, due to e.g. synchronised activity during major (e.g., sports) events \cite{chen2011analysis}.
A well-known simple measure to quantify the coincident electricity consumption between households and moreover to see how this coincidence depends on the number of households is the diversity factor between households \cite{Kersting2007}. To determine the diversity factor between households, we first evaluate the maximum coincident demand, $P_{cd}$ and non-coincident demand, $P_{ncd}$,
respectively in 15 minutes and one day time windows. For this purpose, we sum the maximum power demand of all houses every 15 minutes and then divide the $P_{cd}$ with the sum of the maximum power demand of all houses over the course of a day, i.e.~$P_{ncd}$. 

The diversity factor varies from zero to one, where zero indicates no coincident electricity consumption between households, while a diversity factor equal to one shows strong coincidence. As an example, the diversity factor of the ADRES data set, sampled every 15 minutes for 14 days, is shown Fig.~\ref{fig:divercity_factor}(a). To clearly indicate the interaction between houses during a day (i) the Energy consumption trajectory of all 30 households (in grey); (ii) the maximum coincident demand (in red) and (iii) the non-coincident demand (in purple) between 6:00 to 6:15 (Fig.~\ref{fig:divercity_factor}(b)) and 11:30 to 11:45 (Fig.~\ref{fig:divercity_factor}(c)) for day 13 are shown in Fig.~\ref{fig:divercity_factor}(b) and (c).

Looking at the households trajectories, it is clear that the energy consumption of all houses is low at the beginning of the day and then it is gradually increasing until around 11:00. Thus, the value of the diversity factor, which is the ratio between coincident demand ($P_{cd}$), and the non-coincident demand ($P_{ncd}$) increases during daytime. This proves the interaction between the demand of the households during the day and is the reason why the significant spikes in average energy consumption do not disappear after averaging.

To show the relationship between the value of the diversity factor and the number of households, we calculate the diversity factor for the NOVAREF and ENERA data sets, which are composed of 12 and 70 households, respectively, in the Supplementary Note 2. 
Our results demonstrate that the values of the diversity factor are not zero during a day, even for the 70 ENERA houses and for some time intervals the diversity factor ranges between 0.3 to 0.4 or even larger (maximum 0.6). This indicates that regardless of the number of houses coincident demand will take place during certain time intervals (e.g lunch and dinner time) and spikes during that time will not be averaged out. This will result in a spiky load profile, such as the single day averaged ENERA load profile, see also  Supplementary Note 1.

In the next sections, we introduce a generally applicable methodology for creating residential load profiles and demonstrate its applicability through comparisons with power consumption data sets measured in Germany and Austria, as well as the industry standard load profile. In the next Section (Section \ref{sec:Demand_trend}) we present a methodology to extract the averaged load profile (ALP).
\section{Demand trend: Mode decomposition}
\label{sec:Demand_trend}
\begin{figure}[ht]
\begin{centering}
\includegraphics[width=0.9\columnwidth]{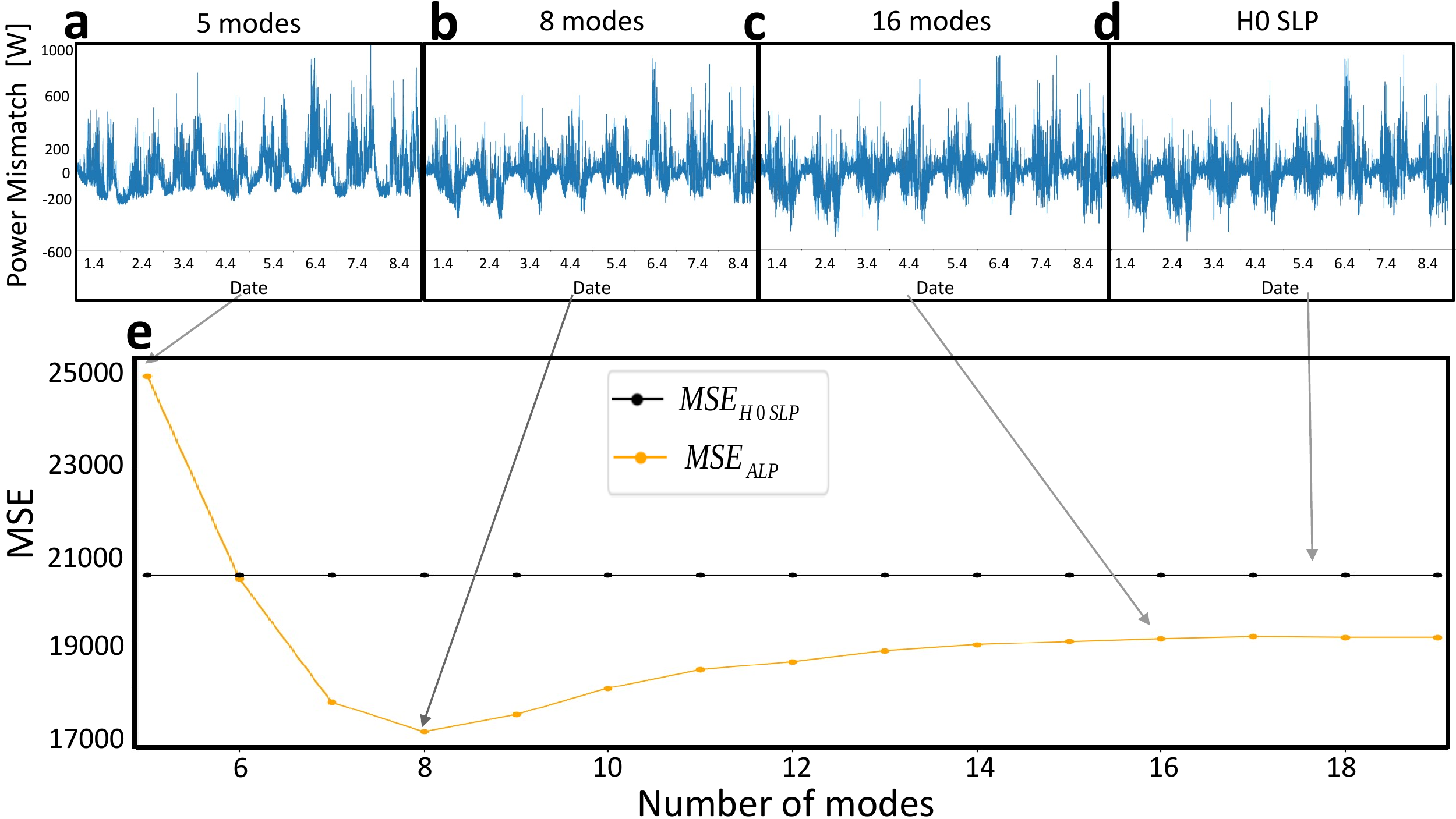}
\par\end{centering}
\caption{\textbf{Determining the optimal number of modes for the average load profile (ALP)}
\textbf{Top panel:} The first three panels from left to right show the $\Delta{P}$ between the original NOVAREF data and the ALP for (a) 5, (b) 8 and (c) 16 summed modes, respectively. The last top panel (d) shows the $\Delta{P}$ between the original NOVAREF data and the H0 SLP. The MSE predictor for the ALP for various $N$ (yellow curve) and the MSE predictor for the H0 SLP (black line) are shown in panel (e). The largest difference between the H0 SLP and the ALP (and so the optimal number of mode to sum) is achieved for $N = 8$, but the ALP clearly overperforms the H0 SLP for any $N \geq 7$.)}
\label{fig:Performance_H0SLP_vs_ALP}
\end{figure}

\begin{figure}[ht]
\begin{centering}
\includegraphics[width=0.9\columnwidth]{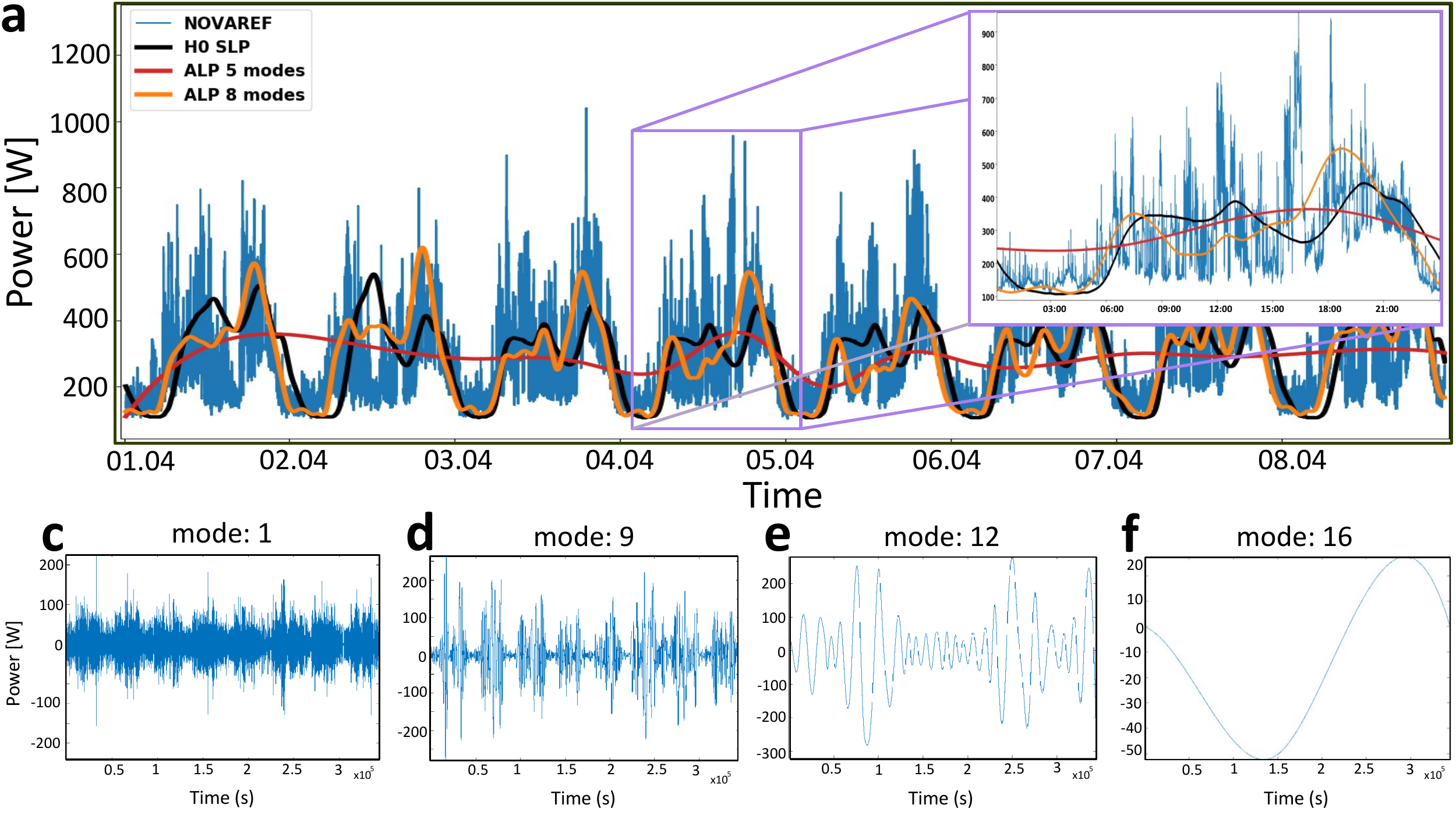}
\par\end{centering}
\caption{
\textbf{The ALP, is a suitable replacement for the H0 SLP with higher precision} (a)-(b) Comparison of the original data with the H0 SLP for the selected dates and the ALP generated through the summing of 5 and 8 modes. The 5 modes sum ALP is not a good approximation for the original data. The 8 modes sum ALP generates a suitable approximation for the signal while excluding the stochastic behaviour of the original data. (c)-(f) Selected examples of the modes extracted using the EMD. We categorise the first nine modes as high modes, and consequently the tenth to eighteenth modes are categorised as low modes. From left to right (c) mode 1, (d) mode 9, (e) mode 12 and (f) mode 16 are shown, respectively.}
\label{fig:Real_data_vs_H0SLP_vs_ALP}
\end{figure}

In this section we present the Average Load Profile predictor, a methodology that extracts the Averaged Load Profile (ALP) from high temporal resolution data sets and creates the load profile of a full week that explicitly respects differences between the different days during a week, and in particular differences between different working days, much as the H0 SLP does. Its main advantage compared to the H0 SLP is that it only requires 4 weeks of high resolution measured data and can, therefore, be applied to both small neighbourhoods and whole cities. Furthermore, its only input is the electricity consumption trajectory and therefore does not rely on numerous micro- or macroscopic parameters, as existing models do. As a result, it can easily be used to analyse both present and future residential power systems, which might include new technologies and devices. Due to the high temporal resolution of the data sets we use, the ALP predictor has a higher level of accuracy than the H0 SLP. It is, therefore, a good alternative to be used instead of the H0 SLP and could provide better and more accurate results, especially when investigating or operating microgrids (a detailed discussion on the subject can be found in Sec.~\ref{sec:Discussion}). In its present form, the ALP predictor can predict the weekly trend of the load profile of a group of houses (anywhere from 12 to 70 or more houses).

To produce the ALP predictor we extract the consumption trend from four consecutive weeks of high resolution electricity consumption data measurements. In this Section we present our results for the NOVAREF data set (temporal resolution = 2 s). 
The methodology we use to create the ALP predictor is the following. We split the measured data into multiple modes using the Empirical Mode Decomposition (EMD) \cite{EEMD}, and so separate the long-term trends from the short-term fluctuations. Because the EMD extracts all the signals present in the data, the original data set can be recreated 1-for-1 without any loss of information by summing up all the modes. A detailed discussion of the adaptive time-frequency data analysis via the EMD can be found in Methods, Supplementary Note 3 and in a recent paper applying EMD in a stochastic context \cite{kampers2020disentangling}.

In order to predict the future daily and weekly household demand we must first train the predictor. As a first step we select 4 chronologically consecutive weeks with no datagaps of measured data from the NOVAREF data set, i.e 07.01--14.01, 14.01--21.01, 04.02--11.02, 11.2--18.04. This constitutes our training set.
Then, we calculate the average of these four weekly consumption profiles and then apply the EMD method to it. Through this process 18 individual and independent modes are extracted. We then divide these modes into two categories: (a) high-frequency; and (b) low-frequency modes. The high-frequency modes contain information on the stochastic consumption fluctuations (see Section \ref{sec:Demand_fluctuations}), while the low-frequency modes are almost free of stochastic fluctuations and are instead dominated by deterministic effects.
Next, we sum the last $N$ low frequency modes.

In order to get the best performing ALP we must determine the optimal number of modes $N$ ($N_{optimal}$). If we sum too few modes, the predictor will fail to generate an ALP that tracks the high resolution consumption better than the H0 SLP. Whereas, if we sum too many modes it will overfit the training data set and lead to inaccurate future load predictions. To determine the $N_{optimal}$ and quantify the performance of the optimised ALP, we use the mean-squared error (MSE) on a validation set, consisting of 29 randomly chosen weeks of data from the NOVAREF data set (see Methods for details on MSE). We must stress here that we used 29 weeks of data because they were available to us. For the purpose of applying this method to other (possibly shorter datasets) only two additional weeks of data are needed to determine the ALP.

The results of this validation analysis reveal the optimal number of modes $N_{optimal}$, which in this case is obtained at $ N_{optimal}= 8 $, see Fig.~\ref{fig:Performance_H0SLP_vs_ALP}. Indeed, even for values of $N>7$, the generated ALP outperforms the H0 SLP in terms of demand prediction. Fig.~\ref{fig:Performance_H0SLP_vs_ALP} shows the results for one of the 29 weeks, and the ALP is found to predict the future daily and weekly electricity consumption of the 12 NOVAREF households much more accurately compared to the H0 SLP. 

Finally, we use 3 randomly chosen weeks to test our predictor and visualise modes and trajectories. We find that the ALP outperforms the H0 SLP when the $N_{optimal}$ is chosen, while it performs worse for a small number of modes, see Fig.~\ref{fig:Real_data_vs_H0SLP_vs_ALP}. Furthermore, we observe the different modes, ranging from high-frequency fluctuations (mode 1) to slowly changing trend (mode 16). We continue to evaluate the ALP performance in the next section (Section \ref{sec:Demand_fluctuations}).

The ALP predictor has several major advantages compared to both the H0 SLP and the available demand side management models. Firstly, it is data driven and is based on modern, high temporal resolution (2 s) measurements, whereas the H0 SLP is based mostly on hourly resolution data measured before 1999, with a small subset of 15-minute resolution data measured between 1990 and 1999 \cite{SLP}. Secondly, because only four chronologically consecutive weeks of data are necessary to generate the ALP and only two additional weeks of data are necessary to determine the $N_{optimal}$, it can be used to predict the electricity consumption of both small and large groups of houses and does not need the plethora of the micro-and-macro-parameters that presently available demand side management models require. Lastly, the ALP has a higher level of accuracy than the H0 SLP due the high temporal resolution of the data sets it uses.

In the next Section (Section  \ref{sec:Demand_fluctuations}) a stochastic model to generate the stochastic fluctuations around the trend is presented.

\section{Demand fluctuations: Stochastic model}
\label{sec:Demand_fluctuations}
\begin{figure}
\begin{centering}
\includegraphics[width=0.9\columnwidth]{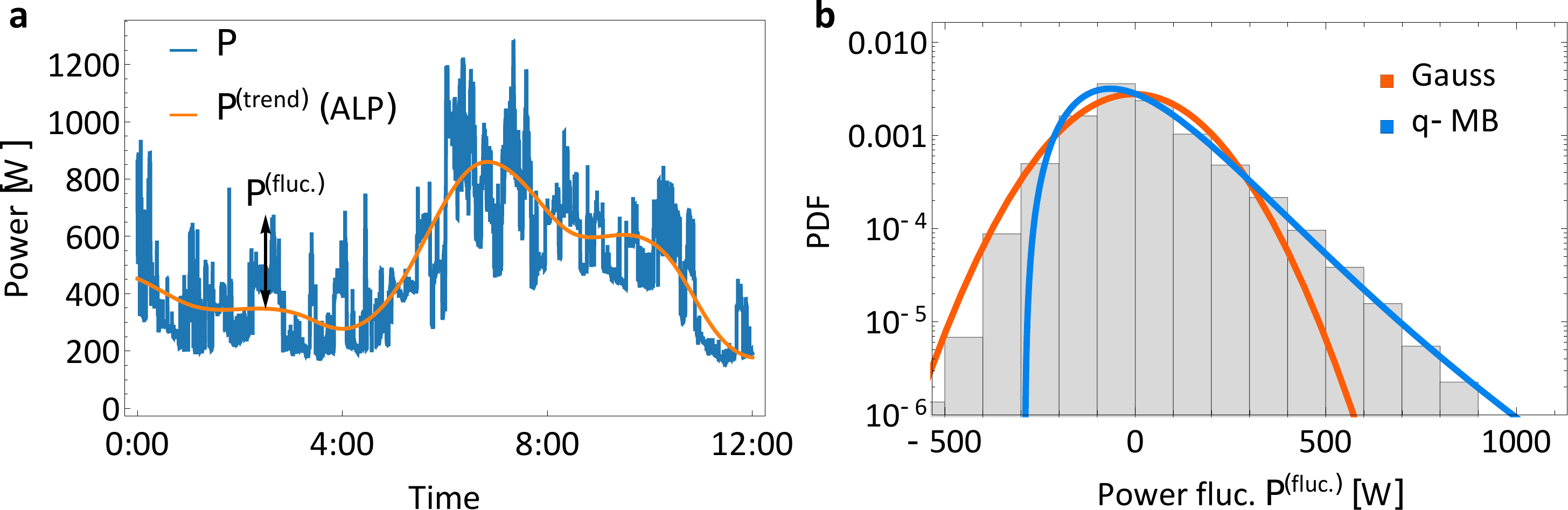}
\par\end{centering}
\caption{
\textbf{The intermittent characteristic of power consumption fluctuations.}
(a) The total power consumption $P$ is a sum of the trend consumption $P^\text{trend}$, obtained by the EMD method described in the previous section and fluctuations $P^\text{fluc.}$. We record the difference between trend and real demand as the fluctuation trajectory.
(b) The probability density function (PDF) of the consumption fluctuation does not follow a Gaussian distribution but is better described by a $q$-Maxwell-Boltzmann distribution, especially on the right flank. The histogram uses the NOVAREF data from 2018, with Gaussian and $q$-Maxwell-Boltzmann parameters determined by the methods of moments. 
\label{fig:DemandFluctuationIntro}}
\end{figure}

\begin{figure}
\begin{centering}
\includegraphics[width=0.9\columnwidth]{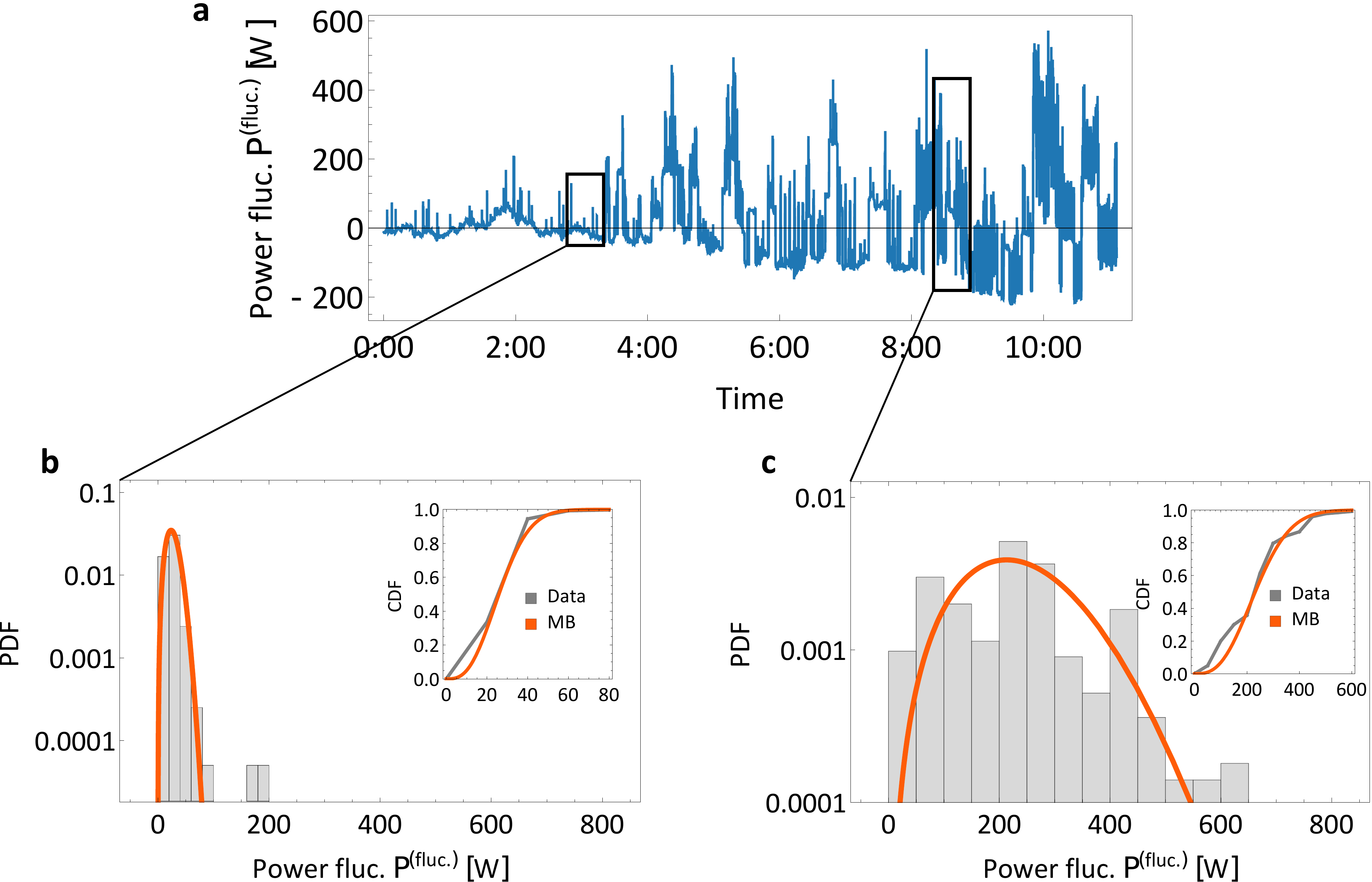}
\par\end{centering}
\caption{
\textbf{Local Maxwell-Boltzmann distributions of power demand fluctuations}
(a) Using only the high-frequency modes from the empirical mode decomposition (EMD), we plot the consumption fluctuations over time.
(b-c) Using superstatistical methods \cite{beck2003superstatistics,beck2005time}, we find a long time scale $T\approx 1000 $ seconds, see Methods, leading to local Maxwell-Boltzmann distributions, that can be a very narrow (b) or broad (c). In addition to the PDF, we compare the cumulative probability functions (CDF) of MB distribution with the data in the inset. All plots use the NOVAREF data from April 2018.
\label{fig:DemandSuperstatistics}}
\end{figure}

\begin{figure}
\begin{centering}
\includegraphics[width=0.95\columnwidth]{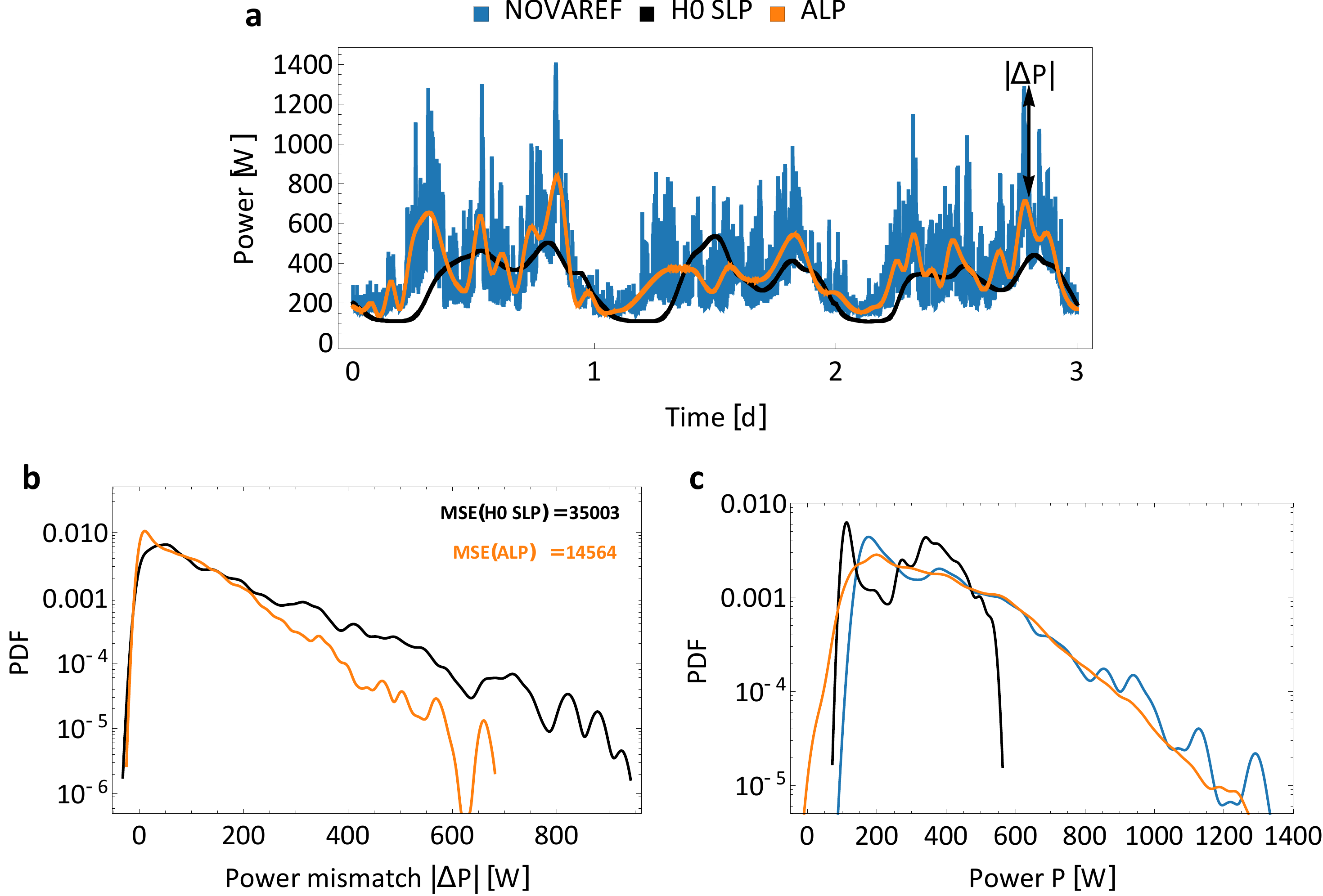}
\par\end{centering}
\caption{
\textbf{Synthetic power demand in agreement with empirical data.}
(a) We display a brief trajectory of the real (blue), H0 SLP (black) and the new ALP (orange). The ALP curve overall keeps closer to the real consumption values than the H0 SLP does.
(b) The histogram of the power mismatch $\Delta P=|P(\text{real})-P(\text{model})|$ shows higher deviations of the demand forecast for the H0 SLP compared to the ALP. 
We also report the mean-squared error (MSE) of the H0 SLP and the ALP with respect to the NOVAREF data set.
(c) Combining the trend extraction via EMD to obtain the ALP and the superstatistical model for demand fluctuations (DLP) approximates the real consumption histogram very well, in particular for large consumption values. All trajectories use the same time stamps, starting 15 minutes past midnight on 15th of April of the NOVAREF data set. The histograms return almost identical results for other weeks.}
\label{fig:DemandSyntheticRealComparison}
\end{figure}

Having extracted the predominantly deterministic trend of the consumption, we now turn to the stochastic fluctuations around this trend. Specifically, we split the power consumption into trend $P^{\text{(trend)}}$ and fluctuations $P^{\text{(fluc.)}}$:
\begin{equation}
P=P^{\text{(trend)}}\left(t\right)+P^{\text{(fluc.)}}\left(t\right).
\end{equation}
First, we investigate the statistical properties of these consumption fluctuations. Analysing the histograms, we notice that the fluctuations are skewed, i.e. asymmetric, and heavy tailed, so large deviations are much more likely than if they were characterised by a simple Gaussian distribution, see Fig.~\ref{fig:DemandFluctuationIntro}.

Next, we construct a stochastic model, which describes the observed consumption fluctuation statistics by applying superstatistics methods  \cite{beck2003superstatistics,beck2005time, weber2018wind, schafer2018non, williams2019superstatistical}: Dividing the full trajectory into several shorter trajectories allows us to characterise each local distribution with a simpler distribution, such as a Gaussian or an exponential distribution. In the case of consumption fluctuations, the fluctuations within each time window of approximately $T\approx 1000$ seconds follow a distinct Maxwell-Boltzmann distribution, see Fig.~\ref{fig:DemandSuperstatistics}. See Supplementary Note 4 for details on the superstatistical procedure such as the extraction of this long time scale $T$. Each local Maxwell-Boltzmann distribution has its distinct scale parameter $\sigma_{MB}$ and offset from zero $\mu_{MB}$. When we now move our analysis from one local distribution to the next one, we observe a slow time dynamics of these Maxwell-Boltzmann parameters $\sigma_{MB}$ and $\mu_{MB}$ and thereby a time dynamics of the local Maxwell-Boltzmann distribution itself. Superimposing these time-varying local distributions, we re-obtain the full aggregated statistics as approximately a q-Maxwell-Boltzmann distribution, see Fig.~\ref{fig:DemandFluctuationIntro}.

Mathematically, we formulate stochastic equations of motion for the fluctuations leading to local Maxwell-Boltzmann distributions as follows:
We define auxiliary variables $x_{i}$, with $i\in\{1,2,...,J\}$, each following a simple Ornstein-Uhlenbeck process, based on independent Wiener processes $W_i$:
\begin{eqnarray}
\text{d}x_{i}\left(t\right)&=&- \gamma x_{i}\left(t\right)\text{d}t+\epsilon \text{d}W_i, 
\end{eqnarray}
with damping $\gamma$ and fluctuations amplitude $\epsilon$.
Hence, the $x_{i}$ are identical but independently distributed Gaussian random variables with a mean 0 
and standard deviation $\sigma=\frac{\epsilon}{\sqrt{2 \gamma}}$. 
Then, the demand fluctuations $P^{\text{(fluc.)}}$ are obtained by aggregating these Gaussian distributions and applying the observed shift from zero $\mu_{MB}$: 
\begin{eqnarray}
P^{\text{(fluc.)}}\left(t\right)&=&
\sqrt{\left(x_{1}(t)\right)^{2}+\left(x_{2}(t)\right)^{2}+...+\left(x_{J}(t)\right)^{2}}+\mu_{MB}.
\end{eqnarray}
As is known from statistical physics \cite{laurendeau2005statistical}, choosing $J=3$ yields exactly Maxwell-Boltzmann distributions in the probability density $p$:
\begin{eqnarray}
p\left(  P^{\text{(fluc.)}}\right)=\frac{1}{\sigma_{MB}^3} \sqrt{\frac{2}{\pi}} \left( P^{\text{(fluc.)}} -\mu_{MB} \right)^2 \exp \left[ -\frac{\left( P^{\text{(fluc.)}} -\mu_{MB} \right)^2}{2 \sigma_{MB}^2} \right],
\end{eqnarray}
where the shape parameter $\sigma_{MB}$ of the local Maxwell-Boltzmann distribution is identical to the standard deviation of the independent Gaussian variables $\sigma_{MB}= \sigma$. We do consider cases of $J\neq 3$ in Supplementary Note 4 and find that $J=3$ is the best fit to the data. Indeed, the approach of 3 independent Gaussian variables is very convenient for the computational application. Alternative modelling approaches using a mathematically simple 1-D process would require more complex dynamics.

In applying the superstatistical approach, we implicitly assumed a separation of time scales. Here, we have the long time scale, on which we locally observe Maxwell-Boltzmann distributions of the power fluctuations of $T\approx 1000$ seconds (Fig. \ref{fig:DemandSuperstatistics}). Furthermore, we estimate the short time scale, on which each local distribution relaxes to its equilibrium, based on the autocorrelation decay of the data as $\tau=1/\gamma\approx 300...400$ seconds, see also Supplementary Note 4. Comparing these two time scales, we observe a clear time separation between long time scales $T$ and short time scales $\tau$, which differ by a factor of 20. Hence, each local Maxwell-Boltzmann distribution relaxes much faster towards its equilibrium (with rate $1/\tau=\gamma$) than the overall process changes towards a new Maxwell-Boltzmann distribution (which happens with a rate of $1/T$).

Finally, we combine the EMD-based trend of the demand with the stochastic fluctuation model, obtaining a data-driven load profile (DLP) and, then, compare it to the original NOVAREF consumption data. We notice that while the precise trajectories are not identical (by construction), the stochastic properties align very well with drastically reduced error compared to the standard H0 SLP model (see Fig.~\ref{fig:DemandSyntheticRealComparison}). In the Methods section we provide a link to the software and pseudo-code to generate these kind of trajectories for other demand regions and data sets.


\section{Discussion}
\label{sec:Discussion}
Summarising, we have shown that modern residential electricity load profiles strongly differ from the H0 standard load profile (H0 SLP), which is widely used in the industry. We set out to replace or supplement the existing standard load profile and obtained three main results:
First, using the empirical mode decomposition (EMD) method, we developed a method to approximate the average load profile (ALP). Second, using superstatistics, we then developed a simple stochastic fluctuation profile (SFP) to describe the non-Gaussian fluctuations of the demand around the trend. Finally, combining both trend and fluctuations yields a full data-driven load profile (DLP).

All three findings are critical for the stable operation of the power system: Knowledge of the expected demand is critical for energy providers to calculate how much power is needed by each household within a given time period. Simultaneously, knowledge of how much the demand might fluctuate around this trend is also essential, to have sufficient balancing and back-up power at hand. 
According to the Bundesverband der Energie- und Wasserwirtschaft e. V. (BDEW) \cite{SLP}, the electricity load of a household can be expected to deviate between 10 \% and 20 \% at any given time. Our fluctuation model provides a better quantitative estimate for these fluctuations to avoid an overestimation of the demand and too much usage of expensive quickly dispatchable generation \cite{wood2013power}. On the other hand, we have to prevent an underestimation of the demand as this could easily lead to a collapse of the system.

Another strong point of our modelling approach is its flexibility: We do not present a fixed load profile but a methodology to extract trend (ALP) and fluctuations (SFP) for any present or future power system. This can be applied to different regions or even continents. Maybe even more importantly, the extracted profiles can easily be updated based on recent developments in consumption, e.g.~due to additional PV installation or adaptation of electrical cars. Similarly, new load profiles can be created for newly built micro or smart grids \cite{fang2011smart}.

The H0 SLP focused mainly on the time scale of 15 minutes and reports much smaller fluctuations than our new load profiles. How can we explain this? First, we note that the H0 SLP used older data, with a temporal resolution much lower than the 1 seconds of our data sources. More relevant is however the power system perspective: Generation and demand are scheduled for fixed time intervals, such as 15 minutes and all fluctuations and deviations within each interval are taken care by control mechanisms \cite{machowski2011power}. 
This view also emphasises the role of the high-voltage transmission grid where some demand fluctuations, which we observe on the distribution grid, might not occur.
Our temporally highly resolved model and focus on the distribution grid becomes increasingly relevant: Conventional generators and their stabilising inertia are removed from the grid and renewable generators are often directly coupled with households. Hence, the balance of supply and demand has to be present also on the distribution level and on an increasingly fast time scale.
Finally, we note that not only do both households and renewable introduce fluctuations in the distribution power grid but their fluctuations share similarities. In particular, the heavy tails of consumption fluctuations and its slowly decreasing power spectrum are also observed in renewable generation \cite{Anvari2016}.

Our model is also especially useful in the case of micro-grids coupled with smart meters. Micro-grids are small autonomous grids, which can function independently or semi-independently from the main power grid \cite{Microgrids_Olivares_et_al_2014}. In the past, small autonomous grids were comprised mainly by remote, usually rural villages or households powered by fossil fuel generators. The concept has been reconfigured to help overcome the challenges of integrating intermittent renewable energy production to the main energy grid \cite{Microgrids_Olivares_et_al_2014}, \cite{Microgrids} as well as provide electricity to populations in remote regions while replacing fossil fuels with renewable resources. Such grids however, always face the problem of matching the intermittent energy generation with the, as we have shown in Section \ref{sec:Necessity_of_new_demand_models}, also intermittent power consumption of the households that comprise them. A task that is made harder by the fact that many microgrids are composed of less than the 332 households that the H0 SLP assumes. As showed in Sections \ref{sec:Necessity_of_new_demand_models} and \ref{sec:Demand_trend}, the consumption behaviour of a small number of houses deviates significantly from the H0 SLP and displays a strong stochastic component. Which technologies can be used to achieve a successful balancing between the generation and consumption \cite{Microgrids_Olivares_et_al_2014}, \cite{Microgrids_management}, which protection schemes \cite{Microgrids_Power_schemes_development} and control systems can be used to ensure the stable and secure operation of these grids  \cite{Microgrids_voltage_regulation}, \cite{Microgrids_Frequency_Inertia}, \cite{Microgrid_operation_after_islanding} and even more scheduling and operation issues \cite{Microgrids_Hirsch}, \cite{Microgrids}, \cite{Micro_grids_planning_operation_and_protection} are all active areas of research.

Furthermore, our model could allow the administrators of an autonomous or semi-autonomous microgrid equipped with smart meters to use the collected data to create an accurate prediction of the electricity consumption of the microgrids units in the following months and take the appropriate measures to ensure a continuous stable operation, e.g. by ensuring enough balancing power is available. At present the H0 SLP, which does not differentiate between the electricity consumption of a handful of houses and a small city, makes this impossible. Our model would also provide researchers, who have access to smart meter data (or any other high temporal resolution data), with much more accurate electricity consumption predictions compared to the ones that H0 SLP allows them at the moment. These prediction could then be used as input for further consumption models. 

Because our approach is entirely dependent on data with no prior assumptions being made, it could theoretically be used to predict the electricity consumption not only of households but also of small businesses or full countries. The current lack of freely available data makes this a task for the future.
For example, we wish to apply our model on consumption data of other regions, such as the UK or the US and compare it to the local equivalent of the H0 SLP. Regions like Canada, where a large number of smart meters are already installed, are particularly promising. While we focused here on household demand, the presented methods should also be applicable to extract the demand trend and fluctuations of industry consumption or combinations of very different consumers. 
Finally, the fluctuation modelling could be extended to include longer-term correlations or match the precise increments in the observed data sets. This would move the study of consumption fluctuations closer to the well-researched state of fluctuations in renewable generation.\\


\section*{Methods}
\label{sec:Methods}
\section*{EMD}
The Empirical mode decomposition (EMD) is a well-known method used for nonlinear, non-stationary data sets. It decomposes the data into a finite number of intrinsic mode functions based on the local properties of the data. Therefore, this method is not restricted to linear or stationary time series, as is the case with other methods, such as Fourier spectral analysis. To determine the empirical mode functions, one applies the following steps on a data set: (1) first the envelope of both local maxima and minima in a data set are defined separately. For instance all local maxima are connected by cubic spline lines for the upper envelope and then the same procedure is repeated for the lower envelope. Consequently, all data is confined between upper and lower envelopes. It is worth mentioning that in this method there is no need for the data to have zero crossings, therefore all values can be just positive or negative. (2) Defining the upper and lower envelopes, their mean value, $m_1$ has to be calculated and then subtracted from the original data, i.e. $X(t)-m_1=h_1$, where $X(t)$ is the original data and $h_1$ is called the first component. (3) In the next step the first component, $h_1$ is converted to the intrinsic mode function (IMF), i.e. $h_1$ should have same number of extrema and zero crossings plus the symmetry of upper and lower envelops around zero (for details see \cite{EEMD}). After finding the IMF, $c_1$, we subtract it from the original data: $X(t)-c_1=r_1$. This $c_1$ is the first mode and contains the shortest period of the original data $X(t)$. (4) Steps (1)-(3) are repeated until $r_n$ becomes a monotonic function, and it becomes impossible to define any envelop for that. If we sum up all IMFs and the residue, we reproduce again the original data, i.e.

\begin{equation}
\centering
    X(t)=\sum_{i=1}^nc_i+r_n;
  \label{EMD}
\end{equation} 

Consequently, we can obtain all the intrinsic oscillation modes of the data set with the EMD method.

\section*{Trend extraction}

In this section we present the adaptive time-frequency data analysis we used to created the ALP. It is based on a normalised version of the one step prediction mean squared error (MSE) \cite{MSE} of an estimator.

In order to produce the ALP, we create a predictor of the future daily and weekly household electricity demand which we then train using four chronologically consecutive weeks of high resolution electricity consumption data measurements which have no data gaps. In this case, weeks 07.01–-14.01, 14.01–-21.01, 04.02–-11.02, 11.2–-18.04 of the NOVAREF data set. For accurate and meaningful results we must use data from the same group of houses whose demand we want to predict. 

The first step in this process is calculating the average of the four weeks of data:
\begin{equation}\label{eq:Emean}
    E_{mean}(t) = \frac{1}{4}\sum_{i=1}^{4} E_{i}(t)
\end{equation}
Where $E_{i}(t)$ is the electricity consumption of each week and i is the number of weeks used in the calculation (i = 4).

For the next step we apply the EMD on the averaged weekly electricity demand profile $E_{mean}$. In the case of the NOVAREF data 18 individual and independent modes are extracted. Finally to calculate the ALP we sum the last N low frequency modes.

\begin{equation}\label{eq:ALP}
    ALP = \sum_{i=1}^{N} M_{i+s}
\end{equation}
where $M_{1+s}$ is highest-frequency mode still to be included. In the case of the NOVAREF data, we have $N=8$ and $s=10$, hence all modes from $i=11$ up to $i=18$ are summed up to obtain the ALP.

To determine the optimal number of low modes to sum ($N_{optimal}$) and quantify the performance of the optimised ALP, we use the mean-squared error (MSE). The MSE measures the average squared difference between the estimated values and the actual value \cite{MSE} and in this paper we calculate it as follows:
\begin{equation}\label{eq:MSE}
    MSE = \frac{1}{L}\sum_{i=1}^{L}\left[C(\text{predicted},t_{i}) - C(\text{actual},t_{i}\right)]^2
\end{equation}
where $C(\text{predicted},t_{i})$ is the predicted consumption timeseries, $C(\text{actual},t_{i})$ is the measured consumption timeseries, $t_{i}$ is the time in seconds, 0 $<$ i $<$ L and L is the length of data used to normalise the ALP. L is 7 days measured in 2 seconds increments
\begin{equation}\label{eq:M}
    M = 7 \ days \ * \ 24 \ hours * \ 60 \ minutes \ * \ \frac{60 \ seconds}{2}
\end{equation}

The MSE is calculated for both the ALP ($MSE_{ALP}$) and the H0 SLP ($MSE_{H0SLP}$) and compared (see Fig.~ \ref{fig:Performance_H0SLP_vs_ALP}) in Section \ref{sec:Demand_trend}. The comparison reveals that:
\begin{equation}\label{eq:MSE_comparsison_results}
    MSE_{ALP}<MSE_{H0SLP}
\end{equation}
for the majority of modes summed. Therefore, the ALP performs better than the H0 SLP and can replace it. 

Regardless of the number of modes summed to create the ALP there is always a minimum which fulfils the $MSE_{ALP} < MSE_{H0SLP}$ condition (see Fig.~\ref{fig:Performance_H0SLP_vs_ALP}). This minimum gives the optimal number of low modes $N$ ($N_{optimal}$) that must be summed to create the optimal ALP for a given set of chronologically consecutive weeks four weeks. This minimum varies with the set of weeks used, though for the NOVAREF data set $N_{optimal} = 8$ for the majority of the weeks investigated. A detailed analysis of the model training, validation and testing can be found in the Supplementary Note 3.

The non-randomness of our results were also verified by calculating the fraction of the $MES_{ALP}$ and the $MSE_{H0SLP}$
\begin{equation}\label{eq:MSE_fraction}
    MSE_{fraction} = \frac{MSE_{ALP}}{MSE_{H0SLP}}
\end{equation}
for the $N \approx N_{optimal}$, $MSE_{fraction}<1$. A more detailed analysis can be found in Supplementary Note 3.\\

\section*{Data availability statement}
The ADRES household consumption data were provided by the TU Wien and are available for research purposes upon request \url{https://www.ea.tuwien.ac.at/projects/adres_concept/EN/}. 
The ENERA household consumption data that support the findings of this study were made available by the DLR Institute for Networked Energy Systems and the EWE, respectively. Restrictions apply to their availability as they were used under license for the current study, and so are not publicly available. The data are, however, available from the authors upon reasonable request and with permission of the DLR Institute for Networked Energy Systems and the EWE. The NOVAREF household consumption data that support the findings of this study are available for download under the CC-BY license together with code on how the EMD and fluctuation analyses were performed on OSF: \url{https://osf.io/yu2dm/?view_only=685370675ca145eb88234031158fc32c}.

\begin{acknowledgments}
The authors would like to thank the NOVAREF, ADRES and ENERA projects for the data that they provided, without which this paper would not have been possible. This project has received funding from the European Union’s Horizon 2020 research and innovation programme under the Marie Sklodowska--Curie grant agreement No 840825, the German Ministry for Education and Research (BMBF) under grant no. 03ET4027A for the DYNAMOS project and no. 03EF3055F for the CoNDyNet2 project.
\end{acknowledgments}

\subsection*{Author contributions}
M.A., E.P., B.S., contributed equally. M.A., E.P., B.S., C.B., H.K. and M. T. conceived and designed the research. E.P. and M.A. detrended the data, E.P developed the deterministic model and B.S., C.B developed the stochastic model. All authors contributed to discussing and interpreting the results and writing the manuscript. 

\subsection*{Competing interests}
The authors declare no competing interests.


\bibliographystyle{naturemag}
\bibliography{main}

\begin{thebibliography}{10}
\expandafter\ifx\csname url\endcsname\relax
  \def\url#1{\texttt{#1}}\fi
\expandafter\ifx\csname urlprefix\endcsname\relax\def\urlprefix{URL }\fi
\providecommand{\bibinfo}[2]{#2}
\providecommand{\eprint}[2][]{\url{#2}}

\bibitem{GrandJeanetal2012}
\bibinfo{author}{Grandjean, A.}, \bibinfo{author}{Adnot, J.} \&
  \bibinfo{author}{Binet, G.}
\newblock \bibinfo{title}{A review and an analysis of the residential electric
  load curve models}.
\newblock \emph{\bibinfo{journal}{Renewable and Sustainable Energy Reviews}}
  \textbf{\bibinfo{volume}{16}}, \bibinfo{pages}{6539 -- 6565}
  (\bibinfo{year}{2012}).

\bibitem{SLP}
\bibinfo{author}{Bitterer, R.} \& \bibinfo{author}{habil. B.~Schieferdecker,
  P.~D.}
\newblock \bibinfo{title}{Repräsentative vdew-lastprofile aktionsplan
  wettbewerb, m-32/99}.
\newblock \bibinfo{type}{Tech. Rep.}, \bibinfo{institution}{VDEW},
  \bibinfo{address}{Stresemannallee 23 D-60596 Frankfurt /M}
  (\bibinfo{year}{2001}).

\bibitem{AustriaSLP}
\bibinfo{author}{E-Control}.
\newblock \bibinfo{title}{Sonstige marktregelnstrom kapitel 6 zählwerte,
  datenformate undstandardisierte lastprofile}.
\newblock \bibinfo{type}{Tech. Rep.}, \bibinfo{institution}{Energie-Control
  Austria für die Regulierung der Elektrizitäts- und Erdgaswirtschaft},
  \bibinfo{address}{Rudolfspl. 13A, 1010 Wien, Austria} (\bibinfo{year}{2019}).
\newblock
  \urlprefix\url{https://www.e-control.at/documents/20903/388512/20150716-SoMa-6-V3-4-clean.pdf/39973f05-a048-425a-957c-46342f0659fa}.

\bibitem{EuropeanEnviromentEnergy2018}
\bibinfo{author}{Schilling, S.}
\newblock \bibinfo{title}{Final energy consumption by sector and fuel.}
\newblock \bibinfo{type}{Tech. Rep.}, \bibinfo{institution}{European Enviroment
  Agency} (\bibinfo{year}{2018}).
\newblock
  \urlprefix\url{https://www.eea.europa.eu/data-and-maps/indicators/final-energy-consumption-by-sector-9/assessment-4}.

\bibitem{GREEND}
\bibinfo{author}{Monacchi, A.} \emph{et~al.}
\newblock \bibinfo{title}{Greend: An energy consumption dataset of households
  in italy and austria}.
\newblock In \emph{\bibinfo{booktitle}{2014 IEEE International Conference on
  Smart Grid Communications (SmartGridComm), Venice}},
  \bibinfo{pages}{511--516} (\bibinfo{year}{2014}).

\bibitem{Wright2007}
\bibinfo{author}{Wright, A.} \& \bibinfo{author}{Firth, S.}
\newblock \bibinfo{title}{The nature of domestic electricity-loads and effects
  of time averaging on statistics and on-site generation calculations}.
\newblock \emph{\bibinfo{journal}{Applied Energy}}
  \textbf{\bibinfo{volume}{84}}, \bibinfo{pages}{389--403}
  (\bibinfo{year}{2007}).

\bibitem{MarszalPomianowskaetal2016}
\bibinfo{author}{Marszal-Pomianowska, A.}, \bibinfo{author}{Heiselberg, P.} \&
  \bibinfo{author}{Larsen, O.~K.}
\newblock \bibinfo{title}{Household electricity demand profiles – a
  high-resolution load model to facilitate modelling of energy flexible
  buildings}.
\newblock \emph{\bibinfo{journal}{Energy}} \textbf{\bibinfo{volume}{103}},
  \bibinfo{pages}{487 -- 501} (\bibinfo{year}{2016}).

\bibitem{liu2011passivity}
\bibinfo{author}{Liu, J.}, \bibinfo{author}{Krogh, B.~H.} \&
  \bibinfo{author}{Ydstie, B.~E.}
\newblock \bibinfo{title}{Passivity-based robust control for power systems
  subject to wind power variability}.
\newblock In \emph{\bibinfo{booktitle}{Proceedings of the 2011 American Control
  Conference}}, \bibinfo{pages}{4149--4154} (\bibinfo{organization}{IEEE},
  \bibinfo{year}{2011}).

\bibitem{Milan2013}
\bibinfo{author}{P.~Milan, J.~P., M.~W{\"a}chter}.
\newblock \bibinfo{title}{Turbulent character of wind energy}.
\newblock \emph{\bibinfo{journal}{{Physical Review Letters}}}
  \textbf{\bibinfo{volume}{110}}, \bibinfo{pages}{138701}
  (\bibinfo{year}{2013}).

\bibitem{Anvari2016}
\bibinfo{author}{Anvari, M.} \emph{et~al.}
\newblock \bibinfo{title}{Short term fluctuations of wind and solar power
  systems}.
\newblock \emph{\bibinfo{journal}{New Journal of Physics}}
  \textbf{\bibinfo{volume}{18}}, \bibinfo{pages}{063027}
  (\bibinfo{year}{2016}).

\bibitem{lopez2015demand}
\bibinfo{author}{L{\'o}pez, M.~A.}, \bibinfo{author}{De~La~Torre, S.},
  \bibinfo{author}{Mart{\'\i}n, S.} \& \bibinfo{author}{Aguado, J.~A.}
\newblock \bibinfo{title}{Demand-side management in smart grid operation
  considering electric vehicles load shifting and vehicle-to-grid support}.
\newblock \emph{\bibinfo{journal}{International Journal of Electrical Power \&
  Energy Systems}} \textbf{\bibinfo{volume}{64}}, \bibinfo{pages}{689--698}
  (\bibinfo{year}{2015}).

\bibitem{logenthiran2012demand}
\bibinfo{author}{Logenthiran, T.}, \bibinfo{author}{Srinivasan, D.} \&
  \bibinfo{author}{Shun, T.~Z.}
\newblock \bibinfo{title}{Demand side management in smart grid using heuristic
  optimization}.
\newblock \emph{\bibinfo{journal}{IEEE transactions on smart grid}}
  \textbf{\bibinfo{volume}{3}}, \bibinfo{pages}{1244--1252}
  (\bibinfo{year}{2012}).

\bibitem{EEMD}
\bibinfo{author}{Wu, Z.} \& \bibinfo{author}{Huang, N.~E.}
\newblock \bibinfo{title}{Ensemble emprical mode decomposition: A
  noise-assisted data analysis method}.
\newblock \emph{\bibinfo{journal}{Advances in Adaptive Data Analysis}}
  \textbf{\bibinfo{volume}{01}}, \bibinfo{pages}{1--41} (\bibinfo{year}{2009}).
\newblock \urlprefix\url{https://doi.org/10.1142/S1793536909000047}.
\newblock \eprint{https://doi.org/10.1142/S1793536909000047}.

\bibitem{Parti1980}
\bibinfo{author}{Parti, M.} \& \bibinfo{author}{C.Parti}.
\newblock \bibinfo{title}{The total and appliance-specific conditional demand
  for electricity in the household sector}.
\newblock \emph{\bibinfo{journal}{Bell Journal of Econimic}}
  \textbf{\bibinfo{volume}{11(1)}}, \bibinfo{pages}{309--21}
  (\bibinfo{year}{1980}).

\bibitem{Aigner1984}
\bibinfo{author}{Aigner, D.}, \bibinfo{author}{Sorooshian, C.} \&
  \bibinfo{author}{Kerwin, P.}
\newblock \bibinfo{title}{Conditional demand analysis for estimating
  residential end-use load profiles}.
\newblock \emph{\bibinfo{journal}{The Energy Journal}}
  \textbf{\bibinfo{volume}{5(3)}}, \bibinfo{pages}{81--97}
  (\bibinfo{year}{1984}).

\bibitem{Capassoetal1994}
\bibinfo{author}{Capasso, A.}, \bibinfo{author}{Grattieri, W.},
  \bibinfo{author}{Lamedica, R.} \& \bibinfo{author}{Prudenzi, A.}
\newblock \bibinfo{title}{A bottom-up approach to residential load modeling}.
\newblock \emph{\bibinfo{journal}{IEEE Transactions on Power Systems}}
  \textbf{\bibinfo{volume}{9}}, \bibinfo{pages}{957--964}
  (\bibinfo{year}{1994}).

\bibitem{PaateroLund2006}
\bibinfo{author}{Paatero, J.~V.} \& \bibinfo{author}{Lund, D.}
\newblock \bibinfo{title}{A model for generating household electricity load
  profiles}.
\newblock \emph{\bibinfo{journal}{International Journal of Energy Research}}
  \textbf{\bibinfo{volume}{30}}, \bibinfo{pages}{273--290}
  (\bibinfo{year}{2006}).
\newblock
  \urlprefix\url{https://onlinelibrary.wiley.com/doi/abs/10.1002/er.1136}.

\bibitem{Train1985}
\bibinfo{author}{Train, K.}, \bibinfo{author}{Herriges, J.} \&
  \bibinfo{author}{Windle, R.}
\newblock \bibinfo{title}{Statistically adjusted engineering models of end use
  load curves}.
\newblock \emph{\bibinfo{journal}{Energy}} \textbf{\bibinfo{volume}{10(10)}},
  \bibinfo{pages}{1103--11} (\bibinfo{year}{1985}).

\bibitem{Estheretal2016}
\bibinfo{author}{Esther, B.~P.} \& \bibinfo{author}{Kumar, K.~S.}
\newblock \bibinfo{title}{A survey on residential demand side management
  architecture, approaches, optimization models and methods}.
\newblock \emph{\bibinfo{journal}{Renewable and Sustainable Energy Reviews}}
  \textbf{\bibinfo{volume}{59}}, \bibinfo{pages}{342 -- 351}
  (\bibinfo{year}{2016}).
\newblock
  \urlprefix\url{http://www.sciencedirect.com/science/article/pii/S1364032115016652}.

\bibitem{BethLoadReview}
\bibinfo{author}{Proedrou, E.}
\newblock \bibinfo{title}{A comprehensive review of residential electricity
  load profile modelling}.
\newblock \emph{\bibinfo{journal}{ArXiv}}  (\bibinfo{year}{2019}).

\bibitem{schafer2018non}
\bibinfo{author}{Sch{\"a}fer, B.}, \bibinfo{author}{Beck, C.},
  \bibinfo{author}{Aihara, K.}, \bibinfo{author}{Witthaut, D.} \&
  \bibinfo{author}{Timme, M.}
\newblock \bibinfo{title}{Non-gaussian power grid frequency fluctuations
  characterized by l{\'e}vy-stable laws and superstatistics}.
\newblock \emph{\bibinfo{journal}{Nature Energy}} \textbf{\bibinfo{volume}{3}},
  \bibinfo{pages}{119} (\bibinfo{year}{2018}).

\bibitem{ADRESReport}
\bibinfo{author}{Einfalt, A.} \emph{et~al.}
\newblock \bibinfo{title}{Energie der zukunft publizierbarer endberich,
  adres-concept}.
\newblock \bibinfo{type}{Tech. Rep.}, \bibinfo{institution}{TU Wien}
  (\bibinfo{year}{2012}).

\bibitem{NOVAREFReport}
\bibinfo{author}{Lange, M.} \& \bibinfo{author}{Zobel, M.}
\newblock \bibinfo{title}{Novaref, erstellung neuer referenzlastprofile zur
  auslegung, dimensionierung und wirtschaftlichkeitsberechnung von
  hausenergieversorgungssystemen}.
\newblock \bibinfo{type}{Tech. Rep.}, \bibinfo{institution}{NEXT|ENERGY},
  \bibinfo{address}{Carl-von-Ossietzky-Strasse 15, 26129 Oldenburg}
  (\bibinfo{year}{2016}).

\bibitem{Destatis2018}
\bibinfo{author}{(Destatis), S.~B.}
\newblock \bibinfo{title}{Wirtschaftsrechnungen einkommens- und
  verbrauchsstichprobe ausstattung privater haushalte mit ausgewählten
  gebrauchsgütern und versicherungen}.
\newblock \bibinfo{type}{Tech. Rep.}, \bibinfo{institution}{Statistisches
  Bundesamt (Destatis)} (\bibinfo{year}{2018}).
\newblock
  \urlprefix\url{https://www.destatis.de/GPStatistik/servlets/MCRFileNodeServlet/DEHeft_derivate_00027268/2152603139004.pdf}.

\bibitem{Braunetal2009}
\bibinfo{author}{Braun, M.}, \bibinfo{author}{Büdenbender, K.},
  \bibinfo{author}{Magnor, D.} \& \bibinfo{author}{Jossen, A.}
\newblock \bibinfo{title}{Photovoltaic self-consumption in germany - using
  lithium-ion storage to increase self-consumed photovoltaic energy}.
\newblock In \emph{\bibinfo{booktitle}{24th European Photovoltaic Solar Energy
  Conference, 21-25 September 2009, Hamburg, Germany}}, \bibinfo{pages}{3121 --
  3127} (\bibinfo{year}{2009}).

\bibitem{Muratori2018}
\bibinfo{author}{Muratori, M.}
\newblock \bibinfo{title}{Impact of uncoordinated plug-in electric vehicle
  charging on residential power demand}.
\newblock \emph{\bibinfo{journal}{Nature Energy}} \textbf{\bibinfo{volume}{3}},
  \bibinfo{pages}{193--201} (\bibinfo{year}{2018}).

\bibitem{chen2011analysis}
\bibinfo{author}{Chen, L.}, \bibinfo{author}{Markham, P.},
  \bibinfo{author}{Chen, C.-f.} \& \bibinfo{author}{Liu, Y.}
\newblock \bibinfo{title}{Analysis of societal event impacts on the power
  system frequency using fnet measurements}.
\newblock In \emph{\bibinfo{booktitle}{2011 IEEE Power and Energy Society
  General Meeting}}, \bibinfo{pages}{1--8} (\bibinfo{organization}{IEEE},
  \bibinfo{year}{2011}).

\bibitem{Kersting2007}
\bibinfo{author}{Kersting, W.~H.}
\newblock \emph{\bibinfo{title}{Distribution System Modeling and Analysis}}
  (\bibinfo{publisher}{CRC Press, Boca Raton, FL}, \bibinfo{year}{2007}).

\bibitem{kampers2020disentangling}
\bibinfo{author}{Kampers, G.} \emph{et~al.}
\newblock \bibinfo{title}{Disentangling stochastic signals superposed on short
  localized oscillations}.
\newblock \emph{\bibinfo{journal}{Physics Letters A}} \bibinfo{pages}{126307}
  (\bibinfo{year}{2020}).

\bibitem{beck2003superstatistics}
\bibinfo{author}{Beck, C.} \& \bibinfo{author}{Cohen, E.~G.}
\newblock \bibinfo{title}{Superstatistics}.
\newblock \emph{\bibinfo{journal}{Physica A: Statistical mechanics and its
  applications}} \textbf{\bibinfo{volume}{322}}, \bibinfo{pages}{267--275}
  (\bibinfo{year}{2003}).

\bibitem{beck2005time}
\bibinfo{author}{Beck, C.}, \bibinfo{author}{Cohen, E.~G.} \&
  \bibinfo{author}{Swinney, H.~L.}
\newblock \bibinfo{title}{From time series to superstatistics}.
\newblock \emph{\bibinfo{journal}{Physical Review E}}
  \textbf{\bibinfo{volume}{72}}, \bibinfo{pages}{056133}
  (\bibinfo{year}{2005}).

\bibitem{weber2018wind}
\bibinfo{author}{Weber, J.} \emph{et~al.}
\newblock \bibinfo{title}{Wind power persistence is characterized by
  superstatistics}.
\newblock \emph{\bibinfo{journal}{arXiv preprint arXiv:1810.06391}}
  (\bibinfo{year}{2018}).

\bibitem{williams2019superstatistical}
\bibinfo{author}{Williams, G.}, \bibinfo{author}{Sch{\"a}fer, B.} \&
  \bibinfo{author}{Beck, C.}
\newblock \bibinfo{title}{Superstatistical approach to air pollution
  statistics}.
\newblock \emph{\bibinfo{journal}{arXiv preprint arXiv:1909.10433}}
  (\bibinfo{year}{2019}).

\bibitem{laurendeau2005statistical}
\bibinfo{author}{Laurendeau, N.~M.}
\newblock \emph{\bibinfo{title}{Statistical thermodynamics: fundamentals and
  applications}} (\bibinfo{publisher}{Cambridge University Press},
  \bibinfo{year}{2005}).

\bibitem{wood2013power}
\bibinfo{author}{Wood, A.~J.}, \bibinfo{author}{Wollenberg, B.~F.} \&
  \bibinfo{author}{Shebl{\'e}, G.~B.}
\newblock \emph{\bibinfo{title}{Power generation, operation, and control}}
  (\bibinfo{publisher}{John Wiley \& Sons}, \bibinfo{year}{2013}).

\bibitem{fang2011smart}
\bibinfo{author}{Fang, X.}, \bibinfo{author}{Misra, S.}, \bibinfo{author}{Xue,
  G.} \& \bibinfo{author}{Yang, D.}
\newblock \bibinfo{title}{Smart grid—the new and improved power grid: A
  survey}.
\newblock \emph{\bibinfo{journal}{IEEE communications surveys \& tutorials}}
  \textbf{\bibinfo{volume}{14}}, \bibinfo{pages}{944--980}
  (\bibinfo{year}{2011}).

\bibitem{machowski2011power}
\bibinfo{author}{Machowski, J.}, \bibinfo{author}{Bialek, J.~W.} \&
  \bibinfo{author}{Bumby, J.}
\newblock \emph{\bibinfo{title}{Power system dynamics: stability and control}}
  (\bibinfo{publisher}{John Wiley \& Sons}, \bibinfo{year}{2011}).

\bibitem{Microgrids_Olivares_et_al_2014}
\bibinfo{author}{{Olivares}, D.~E.} \emph{et~al.}
\newblock \bibinfo{title}{Trends in microgrid control}.
\newblock \emph{\bibinfo{journal}{IEEE Transactions on Smart Grid}}
  \textbf{\bibinfo{volume}{5}}, \bibinfo{pages}{1905--1919}
  (\bibinfo{year}{2014}).

\bibitem{Microgrids}
\bibinfo{author}{Lasseter, R.~H.}
\newblock \bibinfo{title}{Microgrids}.
\newblock In \emph{\bibinfo{booktitle}{2002 IEEE Power Engineering Society
  Winter Meeting. Conference Proceedings (Cat. No.02CH37309)}},
  vol.~\bibinfo{volume}{1}, \bibinfo{pages}{305--308 vol.1}
  (\bibinfo{year}{2002}).

\bibitem{Microgrids_management}
\bibinfo{author}{Katiraei, F.}, \bibinfo{author}{Iravani, R.},
  \bibinfo{author}{Hatziargyriou, N.} \& \bibinfo{author}{Dimeas, A.}
\newblock \bibinfo{title}{Microgrids management}.
\newblock \emph{\bibinfo{journal}{IEEE Power and Energy Magazine}}
  \textbf{\bibinfo{volume}{6}}, \bibinfo{pages}{54--65} (\bibinfo{year}{2008}).

\bibitem{Microgrids_Power_schemes_development}
\bibinfo{author}{{Sortomme}, E.}, \bibinfo{author}{{Venkata}, S.~S.} \&
  \bibinfo{author}{{Mitra}, J.}
\newblock \bibinfo{title}{Microgrid protection using communication-assisted
  digital relays}.
\newblock \emph{\bibinfo{journal}{IEEE Transactions on Power Delivery}}
  \textbf{\bibinfo{volume}{25}}, \bibinfo{pages}{2789--2796}
  (\bibinfo{year}{2010}).

\bibitem{Microgrids_voltage_regulation}
\bibinfo{author}{Farrokhabadi, M.}, \bibinfo{author}{{Cañizares}, C.~A.} \&
  \bibinfo{author}{{Bhattacharya}, K.}
\newblock \bibinfo{title}{Frequency control in isolated/islanded microgrids
  through voltage regulation}.
\newblock \emph{\bibinfo{journal}{IEEE Transactions on Smart Grid}}
  \textbf{\bibinfo{volume}{8}}, \bibinfo{pages}{1185--1194}
  (\bibinfo{year}{2017}).

\bibitem{Microgrids_Frequency_Inertia}
\bibinfo{author}{Delille, G.}, \bibinfo{author}{Francois, B.} \&
  \bibinfo{author}{Malarange, G.}
\newblock \bibinfo{title}{Dynamic frequency control support by energy storage
  to reduce the impact of wind and solar generation on isolated power system's
  inertia}.
\newblock \emph{\bibinfo{journal}{IEEE Transactions on Sustainable Energy}}
  \textbf{\bibinfo{volume}{3}}, \bibinfo{pages}{931--939}
  (\bibinfo{year}{2012}).

\bibitem{Microgrid_operation_after_islanding}
\bibinfo{author}{{Katiraei}, F.}, \bibinfo{author}{{Iravani}, M.~R.} \&
  \bibinfo{author}{{Lehn}, P.~W.}
\newblock \bibinfo{title}{Micro-grid autonomous operation during and subsequent
  to islanding process}.
\newblock \emph{\bibinfo{journal}{IEEE Transactions on Power Delivery}}
  \textbf{\bibinfo{volume}{20}}, \bibinfo{pages}{248--257}
  (\bibinfo{year}{2005}).

\bibitem{Microgrids_Hirsch}
\bibinfo{author}{Hirsch, A.}, \bibinfo{author}{Parag, Y.} \&
  \bibinfo{author}{Guerrero, J.}
\newblock \bibinfo{title}{Microgrids: A review of technologies, key drivers,
  and outstanding issues}.
\newblock \emph{\bibinfo{journal}{Renewable and Sustainable Energy Reviews}}
  \textbf{\bibinfo{volume}{90}}, \bibinfo{pages}{402 -- 411}
  (\bibinfo{year}{2018}).
\newblock
  \urlprefix\url{http://www.sciencedirect.com/science/article/pii/S136403211830128X}.

\bibitem{Micro_grids_planning_operation_and_protection}
\bibinfo{author}{Mumtaz, F.} \& \bibinfo{author}{Bayram, I.~S.}
\newblock \bibinfo{title}{Planning, operation, and protection of microgrids: An
  overview}.
\newblock \emph{\bibinfo{journal}{Energy Procedia}}
  \textbf{\bibinfo{volume}{107}}, \bibinfo{pages}{94 -- 100}
  (\bibinfo{year}{2017}).
\newblock
  \urlprefix\url{http://www.sciencedirect.com/science/article/pii/S187661021631726X}.
\newblock \bibinfo{note}{3rd International Conference on Energy and Environment
  Research, ICEER 2016, 7-11 September 2016, Barcelona, Spain}.

\bibitem{MSE}
\bibinfo{author}{Schelter, B.}, \bibinfo{author}{Winterhalder, M.} \&
  \bibinfo{author}{Timmer, J.}
\newblock \bibinfo{title}{Local and cluster weighted modeling for time series
  prediction}.
\newblock In \emph{\bibinfo{booktitle}{Handbook of Time Series Analysis}}
  (\bibinfo{year}{2006}).

\end{thebibliography}
\end{document}